\documentclass[aps,pra,floatfix,superscriptaddress,twocolumn]{revtex4-2}

\usepackage{graphicx}
\usepackage[english]{babel}
\usepackage{amsmath}
\usepackage{amssymb}
\usepackage{tensor}
\usepackage{xcolor}

\newcommand{\vk}{\mathbf{k}}

\newcommand{\br}{\mathbf{r}}
\newcommand{\be}{\begin{eqnarray}}
\newcommand{\ee}{\end{eqnarray}}
\newcommand{\p}{\partial}

\newcommand{\dc}{c^{\dagger}}

\def\ep#1{\langle #1 \rangle}

\begin{document}

\title{Two-component atomic Fermi superfluid with spin-orbital coupling in thin spherical-shell geometry}

\author{Yan He}
\affiliation{College of Physics, Sichuan University, Chengdu, Sichuan 610064, China}
\email{heyan$_$ctp@scu.edu.cn}

\author{Chih-Chun Chien}
\affiliation{Department of Physics, University of California, Merced, CA 95343, USA.}
\email{cchien5@ucmerced.edu}

\begin{abstract}
We present a theory of two-component atomic Fermi superfluid with tunable pairing interaction in a thin spherical shell subject to spin-orbit coupling (SOC). By incorporating SOC into the Fermi superfluid in the BCS-Bose Einstein condensation (BEC) crossover, we obtain the energy spectrum and equations of state. While the order parameter and chemical potential are suppressed by SOC on the BCS side, the former remains positive but the latter may be pushed to negative values by SOC. Meanwhile, the compressibility exhibits kinks as the pairing interaction or SOC varies, indicating singular behavior of higher-derivatives of the free energy despite the presence of the superfluid order parameter. The minimum of the energy dispersion indicates a decoupling of the energy gap from the order parameter, and the compressibility kinks occur when the energy gap approaches zero. We found the particle-hole mixing in the Fermi superfluid essential for the compressibility kinks since a Fermi gas with the same SOC but no pairing interaction only exhibits complicated dispersions but no singular behavior. Therefore, our results show that a combination of superfluid, SOC, and geometry can lead to interesting phenomena. We also discuss possible experimental realizations and implications.
\end{abstract}

\maketitle

\section{Introduction}
Realizations of ultra-cold atomic clouds in thin spherical shells via the bubble trap in the International Space Station~\cite{Carollo22,Lundblad_2023} or phase-separation structures of multi-component gases in spherical harmonic traps on earth~\cite{PhysRevLett.100.185301,PhysRevLett.97.060403,PhysRevLett.129.243402} have kindled intense research on interacting quantum systems in spherical geometry. While there are plenty theoretical studies of bosonic systems~\cite{SphericalBECPRL19,SphericalSFPRL20,SphericalBECNJP20,PhysRevA.102.043305,PhysRevA.103.053306,PhysRevA.109.013301,PhysRevA.104.033318,PhysRevA.106.013309,PhysRevA.75.013611,PhysRevA.104.063310,PhysRevResearch.4.013122,He2023Soliton,Boegel_2023,PhysRevA.107.023319,PhysRevLett.132.026001,Arazo_2021} addressing the thermodynamics and dynamics of atomic Bose-Einstein condensation (BEC) on a spherical surface (see Refs.~\cite{TononiReview23,TONONI20241} for a review), less works of fermionic systems have been reported. Nevertheless, theory of noninteracting Fermi gases~\cite{FreeFermionShpere} and the BCS-Leggett theory of Fermi superfluid in the BCS-BEC crossover~\cite{He22Sphere} have been generalized to thin spherical-shell geometry, and quantum vortices in such a configuration have been discussed~\cite{He23Vortex}. Moreover, two-component Fermi gases with repulsive interactions on a spherical surface can exhibit different structures with or without rotation~\cite{Yan_RepulsiveSphere}. Due to the Pauli exclusion principle, fermions on a spherical surface tend to pile up toward higher orbital angular momentum states, allowing us to explore interesting phenomena associated with kinetic energy or angular momentum.

Meanwhile, spin-orbit coupling (SOC) has been an intriguing topic since its discovery in relativistic quantum mechanics~\cite{MessiahQM,MQM}. Subsequent developments in condensed matter have led to many interesting phenomena and applications~\cite{WinklerBook,Schaffer_2016}, culminating in the discovery of several topological systems~\cite{Kane_TIRev,Zhang_TIRev,ChiuRMP}. Although typical cold-atom systems are charge neutral and non-relativistic, engineered SOC has been introduced via light-atom interactions~\cite{Lin2011,PhysRevLett.114.125301,Huang2016,doi:10.1126/science.aaf6689,Kolkowitz2017} (see Refs.~\cite{Galitski2013,Zhai_2015} for a review). There have been many theoretical studies of cold-atom systems with SOC in conventional geometries~\cite{PhysRevA.78.023616,PhysRevLett.111.125301,PhysRevA.91.033630,PhysRevLett.112.086401,PhysRevLett.108.235301,PhysRevLett.108.035302,PhysRevLett.111.225301}. Here we consider a two-component Fermi superfluid in a thin spherical shell across the BCS-BEC crossover in the presence of SOC within the BCS-Leggett theory. In our discussion of SOC, the spin refers to the two hyperfine states in the Fermi superfluid while the orbital angular momentum naturally arises on a spherical surface. The energy dispersions of the Fermi superfluid are complicated by the spherical geometry, order parameter, and SOC, resulting in highly variable forms as the chemical potential, pairing strength, and SOC coupling constant change.

We will present several findings of the influence of SOC on the spherical Fermi superfluid in BCS-BEC crossover: (1) The chemical potential on the BCS side is suppressed by SOC, a phenomenon also observable in Fermi gases without pairing interaction subject to the same SOC. (2) The order parameter of the Fermi superfluid no longer determines the energy gap of the energy dispersion in the presence of SOC. We will present numerical evidence where the minimum of the energy dispersion approaches zero while the order parameter remains finite. (3) By evaluating the isothermal compressibility, we observe kinks as the pairing interaction or SOC coupling constant changes. Those kinks hint singular behavior of higher derivatives of the free energy. (4) At the compressibility kinks, the order parameter remains finite and indicates the superfluid survives while the minimum of the energy dispersion plunges toward zero. In contrast, the compressibility of a Fermi gas with SOC but without pairing interaction or a planar Fermi superfluid does not exhibit any kink. (5) The aforementioned phenomena survive when the SOC deviates from the isotropic form and becomes anisotropic.

The rest of the paper is organized as follows. Sec.~\ref{Sec:Theory} presents the theory of two-component atomic Fermi superfluid with SOC across the BCS-BEC crossover on a spherical surface. In addition to the equations of state determining the order parameter and chemical potential, the isothermal compressibility is derived for later analyses. Sec.~\ref{Sec:Result} shows the numerical results of the spherical Fermi superfluid with SOC. The decoupling of the energy gap from the order parameter will be clearly demonstrated. The compressibility exhibits kinks indicating singular behavior of higher derivatives of the free energy, which occur whenever the energy gap approaches zero. Sec.~\ref{Sec:AnisotropicSOC} shows that similar phenomena are observable when the SOC becomes anisotropic. Sec.~\ref{Sec:Implication} discusses possible realizations and measurements of the predictions and their implications. Finally, Sec.~\ref{Sec:Conclusion} concludes our work. Appendix A summarizes the corresponding behavior of a two-component Fermi gas without pairing interaction but subject to the same SOC and spherical-shell geometry to contrast the differences between spherical normal gas and superfluid in the presence of SOC. Appendix B shows the results of Fermi superfluids with SOC on a two-dimensional (2D) plane to contrast the geometric effects.

\section{Theory of Fermi superfluid with SOC on sphere}\label{Sec:Theory}
\subsection{Hamiltonian and thermodynamic potential}
We consider a two-component atomic Fermi gas with attractive contact interaction and SOC confined on a spherical surface.
The fermionic superfluid is described by the BCS-Leggett mean field theory \cite{Fetter_book,Leggett}. Following Ref.~\cite{He22Sphere}, one can normalize the length by the radius $R$ of the sphere and the energy by the lowest single-particle kinetic energy on the sphere, $E_0=\hbar^2/(2M R^2)$, with the fermion mass $M$. The normalized Hamiltonian $H=H_{BCS}/E_0$ becomes~\cite{He22Sphere}
\be
H&=&\int_{S^2}d\br\Big[\sum_{\sigma,\rho}\psi_{\sigma}^{\dag}(\br)\hat{T}_{\sigma\rho}\psi_{\rho}(\br)
-\Delta(\br)\psi^{\dag}_{\uparrow}(\br)\psi^{\dag}_{\downarrow}(\br) \nonumber \\
& &-\Delta^*(\br)\psi_{\downarrow}(\br)\psi_{\uparrow}(\br)\Big]+\frac{\Delta^2}{g}.
\label{eq-BCS}
\ee
Here we introduce the spherical coordinates $\br=(\theta,\phi)$ and the surface element $d\br=\sin\theta d\theta\phi$ for the rescaled unit sphere. $\psi_{\sigma}$ is the fermion operators and the index $\sigma=\uparrow,\downarrow$ labels the spins coming from two hyperfine states of the atoms. 
The order parameter of the superfluid is defined as
\be
\Delta(\br)=g\ep{\psi_{\downarrow}(\br)\psi_{\uparrow}(\br)},
\ee
where $g$ is the pairing coupling constant. Since a spherical surface is periodic in every direction, we will search for solutions with uniform $\Delta$ on the whole spherical surface.

The single-particle contribution includes the SOC term and is given by
\be
\hat{T}_{\sigma\rho}=\Big(-\nabla^2_s-\mu_\sigma\Big)\delta_{\sigma\rho}+2\gamma(\mathbf{L}\cdot\mathbf{S})_{\sigma\rho}.
\ee
Here the dimensionless $\mu_\sigma$ and $\gamma$ are introduced. The spherical Laplacian operator is given by
$
\nabla_s^2
=-\Big(\frac{1}{\sin\theta}\frac{\p}{\p\theta}\sin\theta\frac{\p}{\p\theta}+\frac{1}{\sin^2\theta}\frac{\p^2}{\p^2\phi}\Big)$.
We assume the two components of fermions have equal population and set $\mu_\sigma=\mu$.
The second term in $H$ is the spin-orbital coupling with $\mathbf{L}$ being the orbital angular-momentum operator and $S^a=\sigma^a/2$ denoting the spin of the fermion. Here $\sigma^a$ with $a=1,2,3$ are the three Pauli matrices.
Explicitly, the SOC term in the spin space can be expressed as
\be\label{Eq:IsotropicSOC}
2\gamma\mathbf{L}\cdot\mathbf{S}
=\gamma\left(
         \begin{array}{cc}
           L_z & L_- \\
           L_+ & -L_z
         \end{array}
       \right),
\ee
where $L_{\pm}=L_x\pm i L_y$.

In the superfluid ground state, $\Delta$ is taken to be real, and we search for solutions with uniform $\Delta$ over the whole sphere. An expansion of $\psi_{\sigma}(\br)$ by the spherical harmonics gives
\be
\psi_{\sigma}(\br)=\sum_{l,m}c_{lm\sigma}Y_{lm}(\theta,\phi).
\ee
Then the dimensionless Hamiltonian from Eq.~\eqref{eq-BCS} can be written as
\be
H&=&\frac12\sum_{lm}\Psi_{lm}^\dag\left(
  \begin{array}{cccc}
    \xi_+ & A & 0 & -\Delta\\
    A & \xi_- &\Delta & 0\\
    0 & \Delta & -\xi_- & -A\\
    -\Delta & 0 & -A & -\xi_+
  \end{array}
\right)\Psi_{lm} \nonumber \\
& &+\sum_{l,m}\xi_l.
\label{eq-Hm}
\ee
Here we define $\xi_\pm=\xi_l\pm h$, $\Psi_{lm}=(c_{lm,\uparrow},c_{l,(m+1),\downarrow},\dc_{l,-(m+1),\uparrow},\dc_{l,-m,\downarrow})^T$, and introduce the following notations:
\be
&&\xi_l=l(l+1)-\mu-\frac{\gamma}2,\quad h=\gamma(m+\frac12),\\
&&A=\gamma\sqrt{(l-m)(l+m+1)}.
\ee
For a given $l$ and $m$, the above Hamiltonian in its matrix form can be diagonalized by the Bogoliubov transformation~\cite{Fetter_book} to give rise to four eigenvalues $\pm E_1$ and $\pm E_2$. Thus, two independent quasi-particle dispersions are given by
\be
&&E_1=\Big((\xi_l^2+h^2+A^2+\Delta^2)+2\sqrt{\xi_l^2(A^2+h^2)+A^2\Delta^2}\Big)^{1/2}, \nonumber \\
&&E_2=\Big((\xi_l^2+h^2+A^2+\Delta^2)-2\sqrt{\xi_l^2(A^2+h^2)+A^2\Delta^2}\Big)^{1/2}. \nonumber \\
& &
\ee

In order to study thermodynamic properties, we implement the imaginary-time path-integral formalism and write the partition function as~\cite{Nagaosa_book}
\be
Z&=&\int D\Psi^*_{lm}D\Psi_{lm}\exp\Bigg[-\int_0^{\beta}d\tau \Big(\sum_{l,m}\Psi_{lm}^*\frac{\p}{\p\tau}\Psi_{lm}-H\Big)\Bigg]. \nonumber \\
& &
\ee
Transferring the imaginary time to Matsubara frequency~\cite{Fetter_book} with $k_B = 1$ and integrating out the fermions $\Psi_{lm}$, we find that the thermodynamic potential is given by
\be
\Omega&=&-T\ln Z=-\frac{T}2\sum_{n,l,m}\Bigg(\ln\Big[(i\omega_n)^2-E_1^2\Big] \nonumber \\
& &+\ln\Big[(i\omega_n)^2-E_2^2\Big]\Bigg)
+\sum_{l,m}\xi_l+\frac{\Delta^2}{g}.
\ee
Here $\omega_n=(2n+1)T$ with $n\in\mathbb{Z}$ are the Matsubara frequencies for fermions.

\subsection{Equations of state}
The gap and number equations determines the properties of a Fermi superfluid. They can be inferred from 
\be
\frac{\p\Omega}{\p\Delta}=0,\quad N=-\frac{\p\Omega}{\p\mu}.
\ee
After summing over the Matsubara frequency~\cite{Fetter_book}, we find the  gap and number equations as
\be
&&\frac{2\Delta}g=\frac12\sum_{l,m}\Big(\frac{\p E_1}{\p\Delta}[1-2f(E_1)]
+\frac{\p E_2}{\p\Delta}[1-2f(E_2)]\Big),\\
&&N=\sum_{l,m}\Big(1+\frac12\frac{\p E_1}{\p\mu}[1-2f(E_1)]+\frac12\frac{\p E_2}{\p\mu}[1-2f(E_2)]\Big). \nonumber \\
& &
\ee
Here the derivatives of $E_{1,2}$ are
\be
&&\frac{\p E_1}{\p\Delta}=\frac{\Delta}{E_1}\Big(1+\frac{A^2}{E_0}\Big),\quad
\frac{\p E_2}{\p\Delta}=\frac{\Delta}{E_2}\Big(1-\frac{A^2}{E_0}\Big), \\
&&\frac{\p E_1}{\p\mu}=-\frac{\xi_l}{E_1}\Big(1+\frac{A^2+h^2}{E_0}\Big),~
\frac{\p E_2}{\p\mu}=-\frac{\xi_l}{E_2}\Big(1-\frac{A^2+h^2}{E_0}\Big) \nonumber \\
& &
\ee
with $E_0=\sqrt{\xi_l^2(A^2+h^2)+A^2\Delta^2}$.
As a consistency check, we consider the case without SOC by setting $\gamma=0$. This leads to $A=h=0$ and $\xi_l=l(l+1)-\mu$. We confirm that the above equations recover the results of Ref.~\cite{He22Sphere} for Fermi superfluid on a thin spherical shell in the BCS-BEC crossover.

As a comparison, we construct the Fermi energy of a noninteracting two-component Fermi gas with the same particle number. The ground state of the noninteracting fermions will fill up to the energy level with $l=L_m$, giving the total particle number
$N=2\sum_{l=0}^{L_m}(2l+1)=2(L_m+1)^2$.
Therefore, the Fermi energy and wave vector of the noninteracting Fermi gas are $E_F/E_0= L_m(L_m+1)$ and $k_F=\sqrt{2ME_F}$, which will serve as energy and inverse-length units.

The gap equation of a 2D Fermi superfluid is regularized by introducing the two-body binding energy $\epsilon_b=-1/(Ma^2)$ associated with the two-body $s$-wave scattering length $a$, as discussed in Ref.~\cite{PhysRevLett.115.240401}. For the planar geometry, the pairing coupling constant is related to the scattering length by
$\frac 1g=\int\frac{d^2k}{(2\pi)^2}\frac{1}{2\epsilon_\vk+|\epsilon_b|}$.
The regularization has been adopted by Ref.~\cite{He22Sphere} to study Fermi superfluid in a thin spherical shell. Following a similar argument, the regularization condition for a spherical geometry is modified as
\be\label{Eq:Regularization}
\frac 1g=\sum_l \frac{2l+1}{2\epsilon_l+|\epsilon_b|}.
\ee

\subsection{Isothermal compressibility}
The isothermal compressibility is a thermodynamic quantity which helps determine if a phase transition occurs~\cite{LandauSM}. It is defined as
\be\label{Eq:kappaT}
\kappa_T=-\frac{1}{V}\Big(\frac{\p V}{\p P}\Big)_T=\frac{V}{N^2}\Big(\frac{\p N}{\p \mu}\Big)_T,
\ee
which is associated with the second derivative of the thermodynamic potential. Explicitly,
\be
\Big(\frac{\p N}{\p \mu}\Big)_T=-\frac{\p^2\Omega}{\p\mu^2}-\frac{\p^2\Omega}{\p\mu\p\Delta}\,\frac{\p\Delta}{\p\mu},
\ee
where the derivative $\frac{\p\Delta}{\p\mu}$ can be obtained by differentiating the gap equation with respect to $\mu$:
$\frac{\p^2\Omega}{\p\mu\p \Delta}+\frac{\p^2\Omega}{\p\Delta^2}\,\frac{\p\Delta}{\p\mu}=0$.
In the end, we find that
\be
\Big(\frac{\p N}{\p \mu}\Big)_T=-\frac{\p^2\Omega}{\p\mu^2}+\Big(\frac{\p^2\Omega}{\p\mu\p\Delta}\Big)^2
\Big(\frac{\p^2\Omega}{\p\Delta^2}\Big)^{-1},
\ee
where the second derivatives of $\Omega$ at $T=0$ are given by
\be
&&\frac{\p^2\Omega}{\p\mu^2}=-\frac12\sum_{l,m}\Big(\frac{\p^2 E_1}{\p\mu^2}+\frac{\p^2 E_2}{\p\mu^2}\Big),\\
&&\frac{\p^2\Omega}{\p\mu\p\Delta}=-\frac12\sum_{l,m}\Big(\frac{\p^2 E_1}{\p\mu\p\Delta}+\frac{\p^2 E_2}{\p\mu\p\Delta}\Big),\\
&&\frac{\p^2\Omega}{\p\Delta^2}=\frac{2}{g}-\frac12\sum_{l,m}\Big(\frac{\p^2 E_1}{\p\Delta^2}+\frac{\p^2 E_2}{\p\Delta^2}\Big).
\ee
Here the derivatives of $E_{1,2}$ are
\be
\frac{\p^2 E_{1,2}}{\p\mu^2}&=&\Big(\frac{1}{E_{1,2}}+\frac{\xi_l}{E_{1,2}^2}\frac{\p E_{1,2}}{\p\mu}\Big)
\Big(1\pm\frac{A^2+h^2}{E_0}\Big) \nonumber \\
& &\pm\frac{\xi_l(A^2+h^2)}{E_{1,2}E_0^2}\frac{\p E_0}{\p\mu}, \nonumber \\
\frac{\p^2 E_{1,2}}{\p\mu\p\Delta}&=&\frac{\xi_l}{E_{1,2}^2}\frac{\p E_{1,2}}{\p\Delta}
\Big(1\pm\frac{A^2+h^2}{E_0}\Big)\pm\frac{\xi_l(A^2+h^2)}{E_{1,2}E_0^2}\frac{\p E_0}{\p\Delta}, \nonumber \\
\frac{\p^2 E_{1,2}}{\p\Delta^2}&=&\Big(\frac1{E_{1,2}}-\frac{\Delta}{E_{1,2}^2}\frac{\p E_{1,2}}{\p\Delta}\Big)
\Big(1\pm\frac{A^2}{E_0}\Big)\mp\frac{A^2\Delta}{E_{1,2}E_0^2}\frac{\p E_0}{\p\Delta} \nonumber \\
& &
\ee
with $\dfrac{\p E_0}{\p\mu}=-\dfrac{\xi_l}{E_0}(A^2+h^2)$ and $\dfrac{\p E_0}{\p\Delta}=\dfrac{A^2\Delta}{E_0}$.

In the absence of the SOC term ($\gamma=0$), we find that $E_1=E_2=E_l=\sqrt{\xi_l^2+\Delta^2}$ and
\be
\frac{\p^2 E_l}{\p\mu^2}=\frac{\Delta^2}{E_l^3},~
\frac{\p^2 E_l}{\p\mu\p\Delta}=\frac{\xi_l\Delta}{E_l^3},~
\frac{\p^2 E_l}{\p\Delta^2}=\frac1{E_l}-\frac{\Delta^2}{E_l^3}.
\ee
Then $\p N/\p\mu$ becomes
\be
\frac{\p N}{\p\mu}=\sum_{l,m}\frac{\Delta^2}{E_l^3}+\Big(\sum_{l,m}\frac{\xi_l}{E_l^3}\Big)^2
\Big/\sum_{l,m}\frac{1}{E_l^3},
\label{eq-dn-BCS}
\ee
which recovers the expression of the compressibility shown in Eq. (17) of Ref. \cite{Guo13}.

\section{Results}\label{Sec:Result}
\subsection{BCS-BEC crossover with SOC on sphere}
\begin{figure}
\centering
\includegraphics[width=0.8\columnwidth]{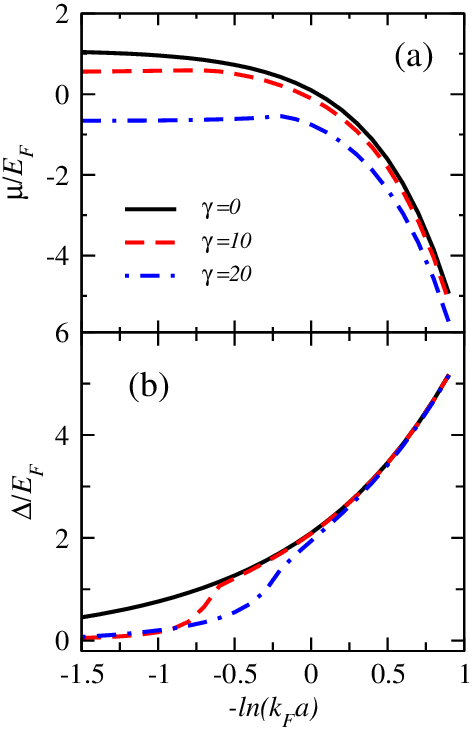}
\caption{(a) Chemical potential $\mu$ and (b) superfluid order parameter $\Delta$ as functions of $-\ln(k_F a)$ on a spherical shell with SOC at $T=0$. All quantities have been normalized. The black, red, and blue lines correspond to $\gamma=0$, $10$, and $20$, respectively. Here $E_F/E_0=110$.}
\label{mu-ka}
\end{figure}

The gap and number equations determine $\mu$ and $\Delta$ once the values of $a$, $T$, and $\gamma$ are given. In the following, we will focus on the ground state at $T=0$ and leave the more complex phenomena at finite temperatures for future research. In Figure \ref{mu-ka}, we plot $\mu$ and $\Delta$ as a function of $-\ln(k_F a)$. For comparison, we choose the SOC coupling constant as $\gamma=0$, $10$ and $20$, respectively. The BCS-BEC crossover behavior characterized by a decrease of the chemical potential along with an increase of the order parameter as the pairing interaction increases survive in the presence of the SOC. 

However, the result shows that there are several effects due to the SOC term. First, it suppresses the chemical potential on the BCS side when compared to the Fermi superfluid without SOC. The suppression of $\mu$ by SOC can be understood by analyzing a two-component Fermi gas without pairing interaction but with the same SOC, as discussed in the Appendix. As the presence of SOC changes the energy dispersions, $\mu$ decreases with $\gamma$ and can even become negative to satisfy the number equation for Fermi gases without pairing interaction. Similarly arguments also apply to $\mu$ of the Fermi superfluid with SOC in the BCS regime. However, the reduction of the chemical potential on the BCS makes it challenging to identify where the BCS-BEC crossover occurs because the conventional definition is when the chemical potential changes signs~\cite{ZwergerBook}. In the presence of SOC, $\mu$ can be negative already on the BCS side. In the following, we will take the point where $\mu=0$ in the absence of SOC, call it the middle of the BCS-BEC crossover, and apply it to the cases with SOC.

Second, the SOC substantially reduces the superfluid order parameter $\Delta$ on the BCS side. As the SOC coupling increases, the suppression of the pairing gap extends beyond the BCS regime. However, $\Delta$ is not completely destroyed, so the ground state remains a Fermi superfluid with SOC in the BCS-BEC crossover. Third, the chemical potential may no longer be a monotonic decreasing function of the pairing interaction in the presence of SOC, this indicating possibly interesting behavior when derivatives with respect to $\mu$ are analyzed. In the following, we will study the isothermal compressibility, which is related to the derivatives of the free energy with respect to $\mu$.

\subsection{Isothermal compressibility and singular behavior}
\begin{figure}
\centering
\includegraphics[width=0.8\columnwidth]{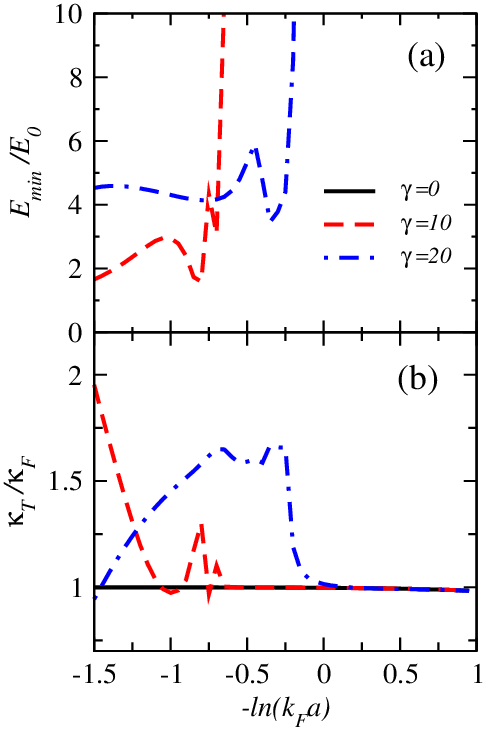}
\caption{(a) Minimum $E_{min}$ of $E_{1,2}$ and (b) isothermal compressibility $\kappa_T$ as functions of $-\ln(k_Fa)$ on a sphere with SOC at $T=0$. The black, red, and blue lines correspond to $\gamma=0$, $10$, and $20$. $\kappa_F$ is the compressibility of a noninteracting two-component Fermi gas with the same total particle number. For $\gamma=0$, $E_{min}=\Delta > 55E_0$ and is not shown in (a).
}
\label{dn-ka}
\end{figure}

One observes in Fig. \ref{mu-ka} kinky behavior of both $\mu$ and $\Delta$ in the presence of SOC. Although $\Delta$ on the BCS side is suppressed by SOC, it still remains non-zero. We notice that in the presence of SOC, $\Delta$ no longer determines the minimum of the energy dispersion which gives rise to the energy gap. To illustrate this point, we plot in Fig. \ref{dn-ka}(a) the minimal energy defined as
\be
E_{min}=\min_{l,m}(E_{1,2}).
\ee
In the absence of SOC and in an Euclidean geometry, the parabolic energy dispersion $\xi_k=k^2/(2M)-\mu$ in $E=\sqrt{\xi_k^2+\Delta^2}$ results in  $E_{min}=\Delta$ if $\mu>0$ and $E_{min}=\sqrt{\mu^2+\Delta^2}$ if $\mu<0$ for Fermi superfluids in the BCS-BEC crossover~\cite{ZwergerBook}. The presence of SOC and the curved space on a sphere alter the energy dispersion in the spectrum and shift the magnitude and location of $E_{min}$ in the Fermi superfluid, thereby causing the decoupling between the order parameter $\Delta$ and the energy gap $E_{min}$. Moreover, one can see that the energy gap given by $E_{min}$ almost vanishes in the presence of SOC on the BCS side but grows on the BEC side back to the value without SOC. However, $E_{min}$ is never truly zero because of avoided crossing due to the finite order parameter $\Delta$. This is in contrast to the Fermi gas with the same SOC but no pairing interaction shown in the Appendix~\ref{FS}, where the energy dispersion crosses zero with no energy gap in the absence of $\Delta$. We also caution that the order parameter $\Delta$ characterizes the symmetry breaking of the internal U(1) symmetry of the Fermi superfluid due to formation of Cooper pairs. Meanwhile, the energy gap $E_{min}$ quantifies the excitation energy to form quasi-particles out of the Fermi superfluid. The decoupling of $E_{min}$ and $\Delta$ in the presence of SOC shows that while the U(1) symmetry remains broken in the Fermi superfluid, the change in the energy spectrum indicates a lower excitation threshold particularly in the BCS regime.

\begin{figure}
\centering
\includegraphics[width=0.8\columnwidth]{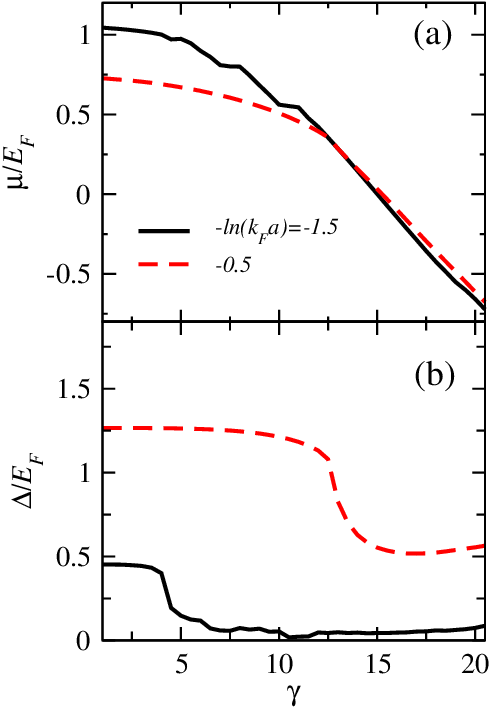}
\caption{(a) $\mu/E_F$ and (b) $\Delta/E_F$ as functions of $\gamma$ on a sphere at $T=0$. The black and red lines correspond to $-\ln(k_F a)=-1.5$ and $-0.5$.}
\label{mu-ga}
\end{figure}

\begin{figure}
\centering
\includegraphics[width=0.8\columnwidth]{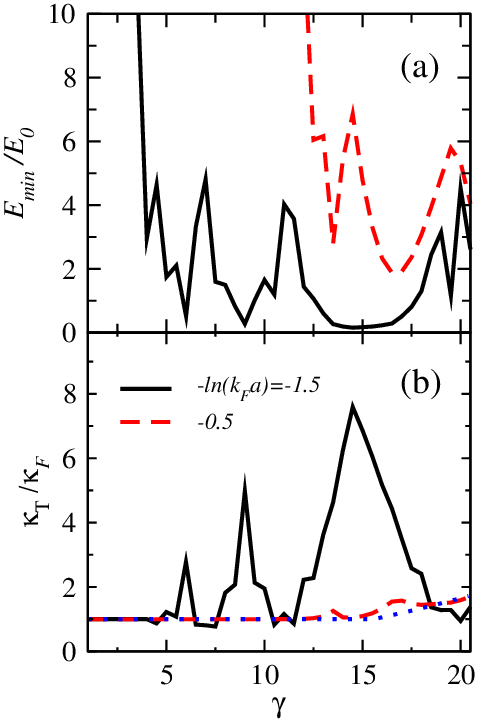}
\caption{(a) The minimum of $E_{1,2}$ and (b) the isothermal compressibility $\kappa_T$ as functions of $\gamma$ on a spherical surface with SOC at $T=0$. The black and red lines correspond to $-\ln(k_F a)=-1.5$ and $-0.5$, respectively. The blue dotted line in (b) shows the result of a Fermi gas with the same SOC but without pairing interaction. $\kappa_F$ is the compressibility of a Fermi gas with the same SOC and total particle number but no pairing interaction.}
\label{dn-ga}
\end{figure}

\begin{figure}
\centering
\includegraphics[width=0.9\columnwidth]{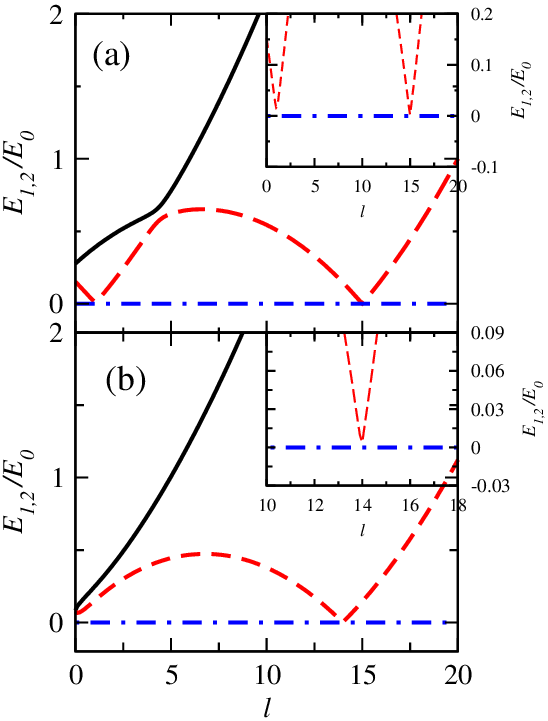}
\caption{Energy dispersion $E_1$ (solid lines) and $E_2$ (dashed lines) as functions of $l$ with $m=0$. Here $-\ln(k_Fa)=-1.5$ and $\gamma=14$ in (a) and $\gamma=15.5$ in (b). The two cases shown here are close to but on the two sides of the highest peak of the black curve in Fig. \ref{dn-ga} (b). The insets show the detail of the dispersion close to zero (indicated by the dot-dash lines), where avoided crossing leads to very small but non-zero minimum in the energy spectrum.
}
\label{E12-L}
\end{figure}

In Fig. \ref{dn-ka}(b), we plot $\kappa_T/\kappa_F$ as a function of $-\ln(k_Fa)$ for three different SOC with $\gamma=0$, $10$ and $20$. Here $\kappa_F$ is the isothermal compressibility of a noninteracting two-component Fermi gas in the absence of SOC on the same sphere with the same total particle number. One can see that if there is no SOC, $\kappa_T$ is a featureless curve that only slightly decreases from the BCS side to the BEC side. In contrast, there are peaks for $\gamma=10$ and $20$ that appear in the shallow BCS regime. The locations of these compressibility peaks correspond to the kinky behavior in the $\mu$ and $\Delta$ curves in Fig. \ref{mu-ka}.

One can see in Fig. \ref{dn-ka}(b)  that $\kappa_T$ may be larger in the BCS limit with negative $-\ln(k_F a)$ than the BEC limit with positive $-\ln(k_F a)$. To understand such asymptotic behavior, we first consider the BCS-BEC crossover without SOC. In this case, the gap and number equations reduce to
\be
\frac1g=\sum_{l,m}\frac{1}{2E_l},\qquad N=\sum_{l,m}\Big(1-\frac{\xi_l}{E_l}\Big)
\ee
with $\xi_l=l(l+1)-\mu$ and $E_l=\sqrt{\xi_l^2+\Delta^2}$.
In the deep BCS limit, $\Delta\approx0$, so we have $E_l\approx|\xi_l|$. Then the number equation can be expressed as
\be
N\approx \int_0^\infty (2l+1)\Big(1-\frac{l^2-\mu}{|l^2-\mu|}\Big)dl=2\mu.
\ee
Thus, we have $\dfrac{\p N}{\p\mu}=2$ in the deep BCS limit, which is in agreement with the asymptotic value without SOC in Fig. \ref{dn-ka}(b) since $\kappa_F=2$ in the same unit. We caution that even though $\Delta$ is negligible, it still causes the energy dispersions to be different from that of a Fermi gas without pairing interaction. This is because the BCS theory includes particle-hole mixing, so the presence of $\Delta$, no matter how small it is, leads to energy spectrum with partially particle-like and partially hole-like features.

Beyond the deep BCS limit, we have to compute $\kappa_T$ according to Eq.~(\ref{eq-dn-BCS}). To get some analytic insights, we approximately convert the summations into integrals as $\sum_{l,m}\to \int (2l)dl$. Then the gap and number equations become
\be
& &\sum_{l,m}\frac{1}{E_l^3}=\frac{1}{\Delta^2}\Big(1+\frac{\mu}{\sqrt{\mu^2+\Delta^2}}\Big),~
\sum_{l,m}\frac{\xi_l}{E_l^3}=\frac{1}{\sqrt{\mu^2+\Delta^2}}. \nonumber \\
& &
\ee
After plugging these equations into Eq.~(\ref{eq-dn-BCS}), we find that $dN/d\mu=2$. This implies that $\kappa_T$ of a Fermi superfluid on a sphere is almost a constant in the whole BCS-BEC crossover, which explains why the result shown in Fig. \ref{dn-ka} without SOC is almost a flat line.

However, the result of the compressibility on a sphere in the BEC limit with SOC is in stark contrast to that of an isotropic 3D Fermi superfluid without SOC, which has the following relation~\cite{Pethick-BEC} 
\be
\mu=-\frac{1}{ma_s^2}+\frac{\pi a_s n}{m}.
\ee
Here $a_s$ is the 3D two-body $s$-wave scattering length. Thus, $\dfrac{dn}{d\mu}=\dfrac{m}{\pi a_s}\to\infty$ as $a_s\to +0$ in the deep BEC limit. The contrast highlights the influences from spatial dimensionality and geometry on interacting quantum systems.

In Fig. \ref{mu-ga}, we plot $\mu$ and $\Delta$ as functions of $\gamma$ for a case on the BCS side with $-\ln(k_F a)=-1.5$ (black lines) and one in the middle of the crossover with $-\ln(k_F a)=-0.5$ (red lines). In panel (a), one can see that $\mu$ decreases from positive to negative in both cases as $\gamma$ increases. Moreover, $\mu$ starts with positive values when $\gamma=0$ but changes signs around $\gamma\approx\sqrt{N}$, which is consistent with the Fermi gas without pairing interaction discussed in Appendix~\ref{App:FS}. In panel (b), one can see that $\Delta$ is suppressed by increasing $\gamma$ in both cases, which is consistent with the suppression of $\Delta$ shown in Fig.~\ref{mu-ka}(b). In the deep-BCS case with $-\ln(k_F a)=-1.5$, $\Delta$ approaches zero as $\gamma$ increases but still remains finite. Therefore, the Fermi gas is still a superfluid in its ground state.

Fig.~\ref{dn-ga} shows the energy gap $E_{min}$ from the energy dispersions and $\kappa_T$ for the same cases shown in Fig. \ref{mu-ga}. We remark that in the presence of SOC, the energy gap is no longer determined by the order parameter $\Delta$, and we numerically scan $E_{1,2}$ to find the minimal value as $E_{min}$. One can see in panel (a) that there are a few places where the energy gap becomes extremely small. However, the presence of avoided crossing due to the small but non-zero $\Delta$ manages to keep the energy gap above zero. As shown in panel (b), there are three peaks of $\kappa_T$ in the deep BCS case shown by the solid curve. The singular behavior of $\kappa_T$ indicates singular higher-derivatives of the free energy as $\gamma$ changes. However, $\Delta$ does not vanish, as shown in Fig.~\ref{mu-ga}, thereby showcasing the non-conventional behavior of a Fermi superfluid on a sphere induced by SOC. We caution that the compressibility peaks in panel (b) result from a complex change of the energy dispersions due to SOC, superfluidity, and spherical geometry. To verify this, the curve of $\kappa_T$ of a Fermi gas with the same SOC but no pairing interaction is shown by the blue dotted line in panel (b), which exhibits smooth behavior without any kinks. The singular behavior of $\kappa_T$ due to SOC and superfluidity on the compact spherical geometry may hint possible structural transitions in energy space without additional symmetry breaking other than the already broken U(1) symmetry due to Cooper-pair formation. Identifying possible energy-space structural transitions induced by SOC, superfluidity, and spherical geometry awaits future research.

To investigate the origin of the singular higher derivatives of the free energy indicated by the kinks of $\kappa_T$ shown in Fig.~\ref{dn-ga}(b), we analyze the energy dispersions $E_{1,2}$ on the two sides of the highest peak in Fig.~\ref{dn-ga}(b) located at $\gamma\approx 15$. Explicitly, Fig.~\ref{E12-L} shows the energy dispersions for $\gamma=14$ and $15.5$, respectively. One can see that the dispersions indeed exhibit different structures: The lower branch approaches zero twice away from $l=0$ when $\gamma=14$ but only once away from $l=0$ when $\gamma=15.5$. The insets of Fig.~\ref{E12-L} emphasize that the energy gap can be extremely small yet still non-zero. Therefore, the structural change in energy space contributes to the suppressed minimum of the energy spectrum, which occurs when $\kappa_T$ exhibits a kink. However, the presence of a small but finite order parameter $\Delta$ plays a significant role in adjusting the energy dispersions and thermodynamic quantities. In Fig.~\ref{E0-L} in Appendix \ref{App:FS}, we show that a two-component Fermi gas without pairing interaction but with the same SOC may exhibit structural changes in its energy dispersions but nevertheless has smooth $\kappa_T$, indicating no singularity in its free energy.

We remark that in our numerical calculations, a cutoff of $L_{\text{cutoff}}=30L_m$ is introduced when carrying out the summations, where $L_m$ is the highest occupied level for a corresponding noninteracting two-component Fermi gas with the same total particle number. As long as the cutoff energy is much larger than the Fermi energy, the numerical results are insensitive to slight variations of the cutoff. This is indeed the case observed in our results, and we have checked that $\mu$ and $\Delta$ are basically the same if $L_{\text{cutoff}}$ is further increased. Moreover, the energy minimum occurs when $l$ is relatively small, as shown in Fig.~\ref{E12-L}. Therefore, increasing the cutoff $L_{\text{cutoff}}$ does not change the location and magnitude of $E_{min}$ and leaves the peaks of $\kappa_T$ intact as well. We also mention that due to the particle-hole mixing in the superfluid when $\Delta$ is finite, although the dispersions of $E_{1,2}$ may look similar to the absolute values of the dispersion $|E_{1,2}^{0}|$ of the Fermi gas without pairing interaction discussed in Appendix \ref{App:FS}, the minimum of the dispersion of the Fermi superfluid does not reach zero due to avoided crossing even if $\Delta$ is suppressed by the presence of SOC.

\section{Anisotropic SOC}\label{Sec:AnisotropicSOC}
After presenting many interesting results with an isotropic form of SOC, here we check if isotropic SOC is necessary by considering an anisotropic SOC term of the form
\be\label{Eq:anisotropicH}
H_{an}&=&2(\gamma_1 L_x S_x+\gamma_2 L_y S_y+\gamma_3 L_z S_z) \nonumber \\
&=&\left(
          \begin{array}{cc}
            \gamma_3 L_z & \gamma_-L_+ +\gamma_+L_- \\
            \gamma_-L_- +\gamma_+L_+ & -\gamma_3 L_z
          \end{array}
        \right)
\ee
with $\gamma_{\pm}=(\gamma_1\pm\gamma_2)/2$. However, the above $H_{SOC}$ cannot be put into a block diagonal form. For a given $l$, therefore, one has to diagonalize a $2(2l+1)$ by $2(2l+1)$ matrix. Nevertheless, it will be greatly simplified if we set $\gamma_1=\gamma_2$. In this case, we have
\be\label{Eq:Hxxy}
H_{SOC}=\left(
          \begin{array}{cc}
            \gamma_3 L_z & \gamma_+L_-  \\
            \gamma_+L_+ & -\gamma_3 L_z
          \end{array}
        \right).
\ee
The BCS Hamiltonian has the same form as shown in Eq.~(\ref{eq-Hm}), but the bare dispersion $\xi_l$ and the constants $A$ and $h$ are modified as follows.
\be
&&\xi_l=l(l+1)-\mu-\frac{\gamma_3}2,\quad h=\gamma_3(m+\frac12),\\
&&A=\gamma_1\sqrt{(l-m)(l+m+1)}.
\ee
After diagonalization, the two independent quasi-particle dispersions are given by
\be
&&E_1=\Big((\xi_l^2+h^2+A^2+\Delta^2)+2\sqrt{\xi_l^2(A^2+h^2)+A^2\Delta^2}\Big)^{1/2},\\
&&E_2=\Big((\xi_l^2+h^2+A^2+\Delta^2)-2\sqrt{\xi_l^2(A^2+h^2)+A^2\Delta^2}\Big)^{1/2}.
\ee

\begin{figure}
\centering
\includegraphics[width=0.7\columnwidth]{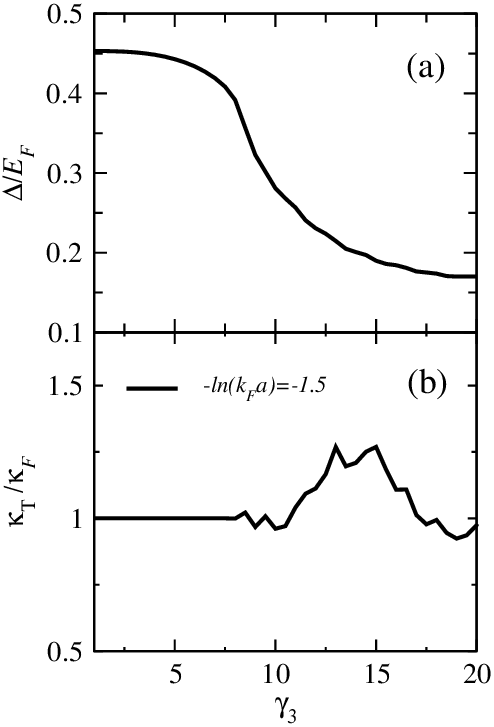}
\caption{(a) Order parameter $\Delta$ and (b) isothermal compressibility $\kappa_T$ of a Fermi superfluid with the anisotropic SOC term shown in Eq.~\eqref{Eq:anisotropicH} as functions of $\gamma_3$ on a spherical surface at $T=0$. Here $\gamma_1=\gamma_2=\gamma_3/2$, $-\ln(k_F a)=-1.5$, and $\kappa_F$ is the compressibility of a Fermi gas with the same SOC and total particle number but no pairing interaction.}
\label{dn-ga-2}
\end{figure}

One may wonder if the kinks of $\kappa_T$ are unique to the isotropic SOC term. To address this question, we solve the number and gap equations with $\gamma_1=\gamma_2=\gamma_3/2$ in the anisotropic SOC term. In the BCS limit with $-\ln(k_F a)=-1.5$ at $T=0$, we plot the superfluid order parameter $\Delta$ and compressibility $\kappa_T$ as functions of $\gamma_3$ in panels (a) and (b) of Fig. \ref{dn-ga-2}, respectively. One can see that there are still some sharp kinks in the $\kappa_T$ curve, but the heights of those peaks are smaller than those of the corresponding isotropic SOC case. Meanwhile, $\Delta$ is also larger than that in the corresponding isotropic SOC case. This means that the smaller $\gamma_{1,2}$ weaken the SOC effects, but the singular behavior of the compressibility still survives when the SOC is no longer isotropic.

Finally, we mention that there are two special cases of Eq.~\eqref{Eq:Hxxy} in which the excitation dispersions will be greatly simplified. The first case is when $\gamma_3=0$, which implies $h=0$.
Then the dispersion becomes
$E_{1,2}=\Big|\sqrt{\xi_l^2+\Delta^2}\pm A\Big|$.
The other case is when $\gamma_1=\gamma_2=0$, which implies $A=0$. Then the dispersion becomes
$E_{1,2}=\Big|\sqrt{\xi_l^2+\Delta^2}\pm h\Big|$.
Although these extreme cases allow simple analytic expressions, they may not exhibit interesting features due to the over-simplifications.

After showing that SOC and superfluidity are crucial for the singular behavior of $\kappa_T$ for the spherical case, we also verify the importance of the compact spherical geometry and the associated orbital angular momentum when the fermions are placed on it. In Appendix~\ref{App:Plane}, we present the results of $\Delta$, $\mu$, and $\kappa_T$ of Fermi superfluids on a 2D plane with a selected type of SOC, which may be considered as the flat-geometry limit of the spherical case. There is no sharp peak in $\kappa_T$ for the planar case, in stark contrast to the spherical results. The comparison between the 2D planar and spherical cases suggests that the singular behavior on a sphere arises from a combination of superfluidity, SOC, and geometry.

\section{Implications}\label{Sec:Implication}
We caution that although singular behavior of the derivatives of the free energy indicates phase transitions in conventional thermodynamic theory~\cite{LandauSM}, the compressibility kinks of Fermi superfluid with SOC on a spherical surface may not fully fit into the picture. First, a thin-shell Fermi superfluid is a finite systems, which is not in the thermodynamic limit. Therefore, singular higher-derivatives of the free energy observed here may not indicate a thermodynamic phase transition due to non-analytic behavior of the free energy in the thermodynamic limit. Second, the superfluid order parameter $\Delta$ remains finite, so there is no further symmetry breaking across the compressibility kinks. Therefore, we may view the compressibility kinks presented here as indicators of different energy spectra of the Fermi superfluid on the two sides of the kink without generating a new phase of matter, which is made possible by the spherical-shell geometry and SOC. 

As Figs.~\ref{dn-ga} and \ref{E12-L} suggest, the difference on the two side of a compressibility kink may lie in the energy dispersions as the pairing strength or SOC varies. Moreover, the energy gap $E_{min}$ may approach zero but never become zero due to avoid crossing as the particle-hole mixing of the superfluid and SOC complicate the energy spectrum. Therefore, the concurrence of the compressibility kink and minimal $E_{min}$ points to an internal transition of the energy spectrum. However, the energy-spectrum transition is not in the conventional sense of thermodynamic phase transitions because the superfluid order parameter remains finite and plays a crucial role in generating the complicated energy dispersion while the higher derivatives of the free energy start to exhibit singular behavior. Experimental realization of such  an internal energy-spectrum transition without symmetry breaking will add to interesting examples of interacting quantum systems in curved space.

Since atomic BEC in a thin spherical shell has been realized using bubble traps or multi-species phase separation~\cite{Carollo22,Lundblad_2023,PhysRevLett.100.185301,PhysRevLett.97.060403,PhysRevLett.129.243402}, similar techniques may apply to fermionic atoms to investigate Fermi gases on a spherical surface. As explained in Ref.~\cite{He22Sphere}, it may take more internal states and confining potentials to trap two-component Fermi gases in a thin spherical shell. After introducing tunable attractive interactions between the fermionic atoms via Feshbach resonance~\cite{Pethick-BEC,Ueda-book}, the ground state is expected to be a Fermi superfluid due to formation of Cooper pairs. Moreover, since a sphere has periodic boundary condition along any direction due to the spherical symmetry, the uniform solution with a constant order parameter $\Delta$ considered here is a natural choice. However, the spherical symmetry may be explicitly broken by imperfections or fluctuations in experiments. To investigate inhomogeneous effects without the spherical symmetry, one may resort to numerical methods like the Bogoliubov-de Gennes equation~\cite{He23Vortex} or Hartree-Fock approximation~\cite{He24_SRepulsive} to explore possible structures of Fermi gases in the presence of SOC on a spherical surface.

For neutral atoms, tunable SOC may be induced by coupling the internal (spin) degrees of freedom of the atoms with their kinetic (orbital) degrees of freedom with the assistance of light-atom coupling~\cite{Lin2011,PhysRevLett.114.125301,Huang2016,doi:10.1126/science.aaf6689,Kolkowitz2017}. Applying such techniques to atoms in a thin spherical shell may be challenging but rewarding because interesting phenomena due to SOC can be investigated in curved space using highly tunable atomic systems. As explained in Sec.~\ref{Sec:AnisotropicSOC}, isotropic SOC is not required for observing the kinks of the compressibility while the superfluid order parameter remains finite. This will relax the requirement in producing suitable SOC terms for observing the singular behavior indicated by the compressibility of atomic Fermi superfluid on a spherical surface.

To verify the decoupling between the superfluid order parameter $\Delta$ and energy gap $E_{min}$ of the excitation in the presence of SOC, one may check quantum-vortex production in a rotating atomic cloud~\cite{Pethick-BEC,He23Vortex} as a signature of superfluidity and also use radio-frequency (rf) spectroscopy~\cite{Kinnuen04,PhysRevLett.101.140403} to measure the excitation gap of the superfluid. On the other hand, the compressibility is related to the sound speed~\cite{Pethick-BEC,Ueda-book}, which is measurable by perturbing the atomic cloud and recording to propagation of the perturbation. With the rapid development of trapping and manipulating cold atoms, one may envision that realization and measurement of Fermi superfluid with SOC on a spherical surface will be within reach of future research.

We mention that the presence of SOC can lead to very different behavior between Fermi gases with and without pairing interaction in a thin spherical shell by contrasting the aforementioned results and those in Appendix \ref{App:FS}. Instead of destroying the superfluid, SOC modifies the particle-hole mixed excitation spectrum and causes singular behavior in the free energy, which can be inferred by the kinks of the compressibility. Even without SOC, the compressibility of Fermi superfluid in 3D drastically differs from that on a spherical surface due to different dimensionality and geometry. Therefore, atomic Fermi superfluid in a thin spherical shell provides a playground for integrating superfluid, SOC, and geometric effects.

\section{Conclusion}\label{Sec:Conclusion}
We have developed the theoretical foundation for describing atomic Fermi superfluid with SOC in a thin spherical shell. While the superfluid order parameter remains finite in the BCS-BEC crossover, the presence of SOC suppresses the chemical potential on the BCS side and introduces complicated energy dispersions. Importantly, the energy gap according to the dispersion with SOC is no longer simply determined by the order parameter. As a consequence, the compressibility exhibits kinks and indicates singular higher-derivatives of the free energy, which reflect internal transitions of the energy spectrum. The survival of the order parameter is crucial because it retains particle-hole mixing and introduces avoided crossing, and we found the compressibility kinks occur when the energy gap approaches zero. Our results thus illustrates interesting interplays between superfluidity, SOC, and the underlying geometry. Future research on realizing atomic Fermi superfluid with SOC on a spherical surface inspired by this work will unveil more exciting physics in cold-atom systems with interesting interactions and geometries.

\begin{acknowledgments} 
Y. H. was supported by the National Key R$\&$D Program of China (No. 2024YFF0508503). C. C. C. was partly supported by the NSF (No. PHY-2310656) and DOE (No. DE-SC0025809).
\end{acknowledgments}

\appendix
\section{Fermi gases with only SOC on sphere}\label{App:FS}
We consider a two-component Fermi gas with the same isotropic SOC shown in Eq.~\eqref{Eq:IsotropicSOC} but no pairing interaction  on a spherical surface. The two energy dispersions become
\be
E_{1,2}^{(0)}=\xi_l\pm\sqrt{A^2+h^2}=\xi_l\pm\gamma(l+1/2).
\ee
Explicitly, they can be expressed as
\be
&&E_1^{(0)}=l(l+1)-\mu+\gamma l,\\
&&E_2^{(0)}=l(l+1)-\mu-\gamma(l+1).
\ee
For large $l$, they can be approximated as
\be
E_{1,2}^{(0)}\approx l^2-\mu\pm\gamma l.
\ee
For the dispersion $E_2^{(0)}$, one can see that the minimum energy occurs at $l\approx\gamma/2$. The number equation becomes
\be
N=\sum_l(2l+1)\Big[\theta(-E_1)+\theta(-E_2)\Big].
\ee
Here $\theta(x)$ is the unit step function. For convenience, we can approximate the above summations by integrals.

\begin{figure}
\centering
\includegraphics[width=0.8\columnwidth]{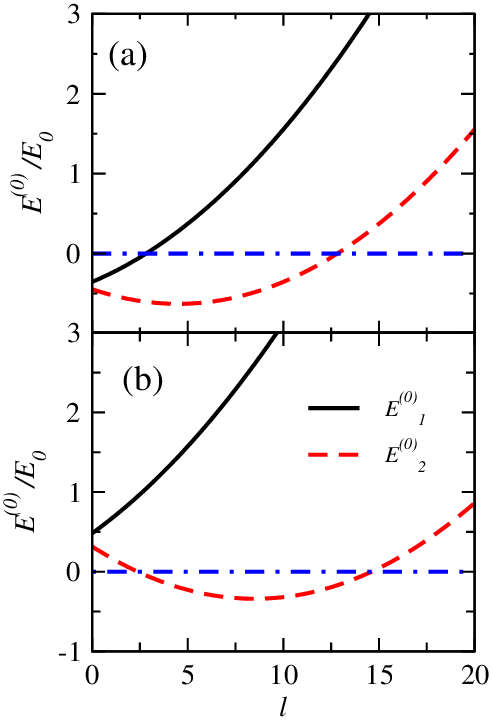}
\caption{Energy dispersion of a two-component Fermi gas with isotropic SOC but no pairing interaction on a spherical surface, showing $E_1^{(0)}$ (solid lines) and $E_2^{(0)}$ (dashed lines) as functions of $l$ for (a) $\gamma=10$ with $\mu>0$ and (b) $\gamma=18$ with $\mu<0$.}
\label{E0-L}
\end{figure}

For $\mu>0$, the dispersions are shown in the top panel of Fig. \ref{E0-L}. The Fermi surface is determined by the positive roots of equations $E_{1,2}^{(0)}=0$ are $L_{1,2}=\dfrac{\sqrt{\gamma^2+4\mu}\pm\gamma}2$. Thus we find that
\be
N=\int_0^{L_1}dl(2l)+\int_0^{L_2}dl(2l)=L_1^2+L_2^2=2\mu+\gamma^2.\nonumber\\
\ee
Therefore, if $\gamma^2>N$, there is no positive-$\mu$ solution.

For $\mu<0$, on the other hand, we always have $E_1^{(0)}>0$. The positive roots of the equation $E_2^{(0)}=0$ are $L^{(2)}_{1,2}=\dfrac{\gamma\pm\sqrt{\gamma^2+4\mu}}{2}$. This situation is shown in the bottom of Fig. \ref{E0-L}. Thus, we find that
\be
N=\int_{L^{(2)}_2}^{L^{(2)}_1}dl(2l)
=(L^{(2)}_1)^2-(L^{(2)}_2)^2=\gamma\sqrt{\gamma^2+4\mu}.\nonumber\\
\ee
Consequently, we find a solution $\mu=\dfrac{N^2-\gamma^4}{4\gamma^2}<0$ if $\gamma^2>N$.
In summary, when $\gamma^2>N$, we have $\mu<0$ for the Fermi gas with only SOC. For a Fermi superfluid in the BCS limit, there will be a small $\Delta$, but we still expect that $\mu<0$ in this case.

For the Fermi gas with only SOC, the number susceptibility is given by
\be
\frac{dN}{d\mu}=\left\{
                  \begin{array}{ll}
                    2, & \gamma^2<N, \\
                    \dfrac{2\gamma}{\sqrt{\gamma^2+4\mu}}=\dfrac{2\gamma^2}{N}, & \gamma^2>N.
                  \end{array}
                \right.
\ee
The corresponding compressibility is shown by the dotted line in Fig.~\ref{dn-ga}. There is no singular behavior in the compressibility without pairing interaction as the strength of SOC varies. The difference between the Fermi gases with and without pairing interaction shows the importance of particle-hole mixing in the energy dispersions, which gives rise to interesting phenomena in superfluid only.

\section{2D planar Fermi superfluid with SOC}\label{App:Plane}
Fermi superfluids in three-dimensional Euclidean space with various types of spin-orbital couplings have been discussed in Ref.~\cite{MeloSOC}, and no singular behavior of the compressibility was reported. Here we consider Rashba-type SOC on a 2D plane with the Hamiltonian 
\be
H_{SOC}=\gamma(k_x\sigma_y-k_y\sigma_x).
\ee
Here $k_x$ and $k_y$ are 2D the wave vectors and $\sigma_a$ with $a=x,y,z$ are the three Pauli matrices. This SOC term may be thought of as the planar limit of the spherical SOC term with the form  $(\br\times\vk)\cdot\mathbf{S}$ with $S_a=\sigma_a/2$.
The Hamiltonian of a 2D planar Fermi superfluid with Rashba-type SOC in momentum space can be written as
\begin{widetext}
\be
&&H=\frac12\sum_{\vk}\Psi_{\vk}^\dag\left(
  \begin{array}{cccc}
    \xi_\vk & \gamma(-k_y-i k_x) & 0 & -\Delta\\
    \gamma(-k_y+i k_x) & \xi_\vk-h &\Delta & 0\\
    0 & \Delta & -\xi_\vk & \gamma(k_y-i k_x)\\
    -\Delta & 0 & \gamma(k_y+i k_x) & -\xi_\vk
  \end{array}
\right)\Psi_{\vk}+\sum_{\vk}\xi_\vk.
\label{eq-Hm}
\ee
\end{widetext}
Here $\Psi_{\vk}=(c_{\vk,\uparrow},c_{\vk,\downarrow},\dc_{-\vk,\uparrow},\dc_{-\vk,\downarrow})^T$ and $\xi_\vk=k^2/(2M)-\mu$ with $\vk=(k_x,k_y)$.

\begin{figure}
	\centering
	\includegraphics[width=0.7\columnwidth]{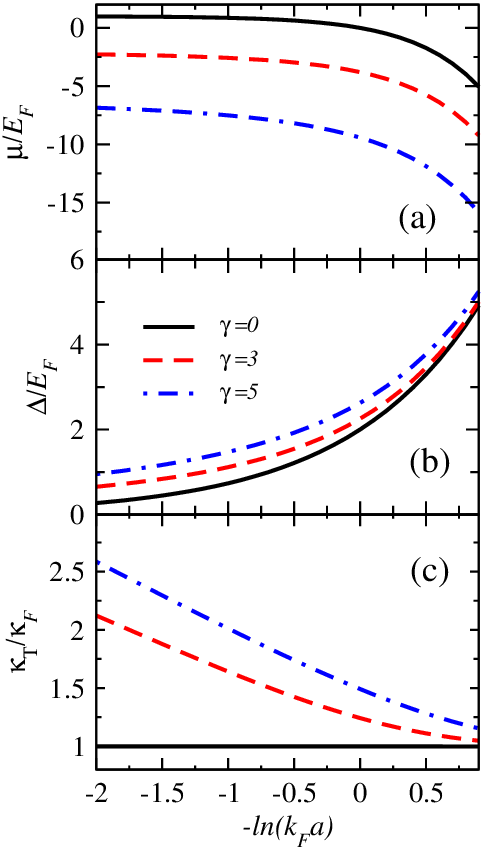}
	\caption{Normalized chemical potential $\mu$ (left), order parameter $\Delta$ (right), and compressibility $\kappa_T$ as functions of $-\ln(k_F a)$ of 2D planar Fermi superfluids with Rashba-type SOC at $T=0$. The solid, dash, and dot-dash lines correspond to $\gamma=0$, $3$, and $5$, respectively.
	}
	\label{dn-2d}
\end{figure}

For a given $\vk$, the above 4 by 4 matrix can be diagonalized to give rise to four eigenvalues $\pm E_1(\vk)$ and $\pm E_2(\vk)$. Thus, there are two independent quasi-particle dispersions:
\be
E_1=\sqrt{\epsilon_1(\vk)^2+\Delta^2},~ E_2=\sqrt{\epsilon_2(\vk)^2+\Delta^2},
\ee
Here $\epsilon_1(\vk)=\xi_\vk+\gamma\sqrt{k_x^2+k_y^2}$ and $\epsilon_2(\vk)=\xi_\vk-\gamma\sqrt{k_x^2+k_y^2}$.
The gap and number equations of the 2D Fermi superfluid with Rashba-type SOC at $T=0$ are given by
\be
&&\frac{1}g=\frac14\sum_\vk\Big(\frac{1}{E_1}+\frac1{E_2}\Big),\\
&&N=\sum_{\vk}\Big[1-\frac12\frac{\epsilon_1(\vk)}{E_1}-\frac12\frac{\epsilon_2(\vk)}{E_2}\Big].
\ee
For a special case without SOC by setting $\gamma=0$, the two quasi-particle dispersions become $E_1=E_2=E_\vk=\sqrt{\xi_l^2+\Delta^2}$, which recover the conventional results without SOC. Moreover, the gap equation can be regularized by the method discussed above Eq.~\eqref{Eq:Regularization}.
The compressibility is defined in Eq.~\eqref{Eq:kappaT}.
Following similar steps as before, the derivative $\p N/\p\mu$ can be computed as 
\be
\frac{\p N}{\p\mu}=\Delta^2\mathcal{B}+\frac{\mathcal{A}^2}{\mathcal{B}}.
\ee
Here $\mathcal{A}=\sum_{\vk}\Big[\frac{\epsilon_1(\vk)}{2E_1^3}+\frac{\epsilon_2(\vk)}{2E_2^3}\Big]$ and $\mathcal{B}=\sum_{\vk}\Big(\frac{1}{2E_l^3}+\frac{1}{2E_2^3}\Big)$.

Figure \ref{dn-2d} shows $\mu$, $\Delta$, and $\kappa_T$ as functions of $-\ln(k_F a)$ at $T=0$ for  $\gamma=0$, $3$, and $5$, respectively. Note that $\kappa_T$ is normalized by $\kappa_F=1/(2\pi)$, which is the compressibility of a 2D noninteracting Fermi gas. The curves of $\mu$ and $\Delta$ have similar trends as the spherical case with isotropic SOC shown in Fig.~\ref{mu-ka}, but there is no kink in the 2D planar case. Furthermore, there is no sharp peak in $\kappa_T$ for the planar case. The contrast suggests that the sharp peaks in $\kappa_T$ of the spherical case has their origin in the compact geometry and the associated orbital angular momentum.

%

\end{document}